\documentclass{article}

\title{Speaker Verification Using Simple Temporal Features and Pitch Synchronous Cepstral Coefficients}
\date{}
\author{Bhavana V. S \and  Pradip K. Das \\Department of Computer Science and Engineering\\Indian Institute of Technology Guwahati, Assam, India}

\usepackage{graphicx}
\usepackage{ragged2e}
\usepackage{float}
\usepackage[square,numbers]{natbib}

\graphicspath{ {./png} }
\begin{document}
\maketitle

\section*{ABSTRACT}
\begin{justify}
Speaker verification is the process by which a speaker’s claim of identity is tested against a claimed speaker by his/her voice. Speaker verification is done by the use of some parameters (features) from the speaker’s voice which can be used to differentiate among many speakers. The efficiency of speaker verification system mainly depends on the feature set providing high inter-speaker variability and low intra-speaker variability. There are many methods used for speaker verification. Some systems use Mel Frequency Cepstral Coefficients as features (MFCCs), while others use Hidden Markov Models (HMM) based speaker recognition, Support Vector Machines (SVM), GMMs, etc. In this paper simple intra-pitch temporal information in conjunction with pitch synchronous cepstral coefficients forms the feature set. The distinct feature of a speaker is determined from the steady state part of five cardinal spoken English vowels. The performance was found to be average when these features were used independently. But very encouraging results were observed when both features were combined to form a decision for speaker verification. For a database of twenty speakers of 100 utterances per speaker, an accuracy of 91.04\% has been observed. The analysis of speakers whose recognition was incorrect is conducted and discussed .
\end{justify}

\section{INTRODUCTION}
\begin{justify}
Speech is one of the primary modes of communication. Human ears can perceive even the minute details of delivered speech and thus distinguish a speaker from another. In a speaker recognition system, the main role is played by the feature set which can distinguish different speakers efficiently. \par In paper \cite{Dormanetal1980} some acoustic cues has been determined to identify the fricative-affricate contrast in word final position. The features used were the temporal and spectral characteristics of the vocalic interval, duration of a silent interval, presence or absence of a release burst, rise time of the fricative noise and duration of the fricative noise. It was reported that onset characteristics of the fricative noise play a significant role in perception of fricative-affricate contrast. At short closure intervals the stimuli were heard as “dish” while at long as “ditch” (the words used were “dish” and “ditch”).\par Paper \cite{Poeppeletal2008} focused on the various activities in the human body during the perception of speech signals into messages. This paper investigated how the raw speech signal is perceived as meaningful data and how the human brain analyse this data. It also looked into the importance of the acoustic properties of the signal in identifying a spoken utterance and also how similar words are distinguished by concentrating on the local areas of the signal. It discussed about the distinct features which influenced the phonological processes and articulatory gestures.\par A representation of the phase information in speech model was presented in \cite{Hernaezetal2009}. Phase is defined as the fractional part of a period through which the time variable of a periodic quantity has moved as measured at any point in time from an arbitrary time origin. Usually, time origin is the last point at which amplitude value passed through a zero position from negative to positive direction. Phase information has been used in speech representation and successful reconstruction of the original signal from its phase values have been reported. It was also reported that harmonic amplitudes play a critical role in the signal as they reflect the spectral energy which contain the basic information about the signal.\par There is a report on speaker characterization by using prosodic super vectors with 'negative within class covariance normalization' projection and speaker modeling with support vector regression in \cite{Longetal2009}. They have also proposed a segmental weight fusion technique combining acoustic and prosodic subsystems. They used SRE corpus and they have reported the error rate of 4.5\% for fusion system.\par Speaker clustering was done in GMM based speaker recognition system \cite{Sunetal2003} in order to reduce computational complexity. ISODATA algorithm was used for clustering speakers whose acoustic characteristics were similar with a distance measure. The database used was that of China National High Technology Project. The identification was 98.75\% correct. Speech was recognized even in the situation where voice characteristics of input speech were unknown \cite{Suzukietal2003}. ASJ-PB and ASJ-JNAS databases were used for the experiments.\par Standard speech/non speech HMM's were conditioned on speaker traits and evaluated on cepstral and pitch features in \cite{Shafranetal2003}. SVMs were applied to speech lattices. Acoustic features and word sequences are used to determine speaker traits. The overall sound characteristics for speakers can be covered by a set of Acoustic Segment Models (ASM). The mean vectors of ASM based on unsupervised MAP adaptation are concatenated to represent characteristics of specific speaker as proposed in \cite{Binetal2009}. The acoustic feature vectors used are 12 MFCCs and normalized energy plus their first and second order time derivatives. Speaker specific characteristics were obtained by the MFCCs. In addition, an objective function consisting of contrastive cost in terms of speaker similarity and dissimilarity as well as data reconstruction cost was used as regularization to normalize non speaker related information \cite{Salmanetal2011}.\par MFCC, due to structure of its filter bank, captures vocal tract characteristics more effectively in lower frequency regions. A new set of features using a complementary filter bank structure which improved distinguishing ability of speaker specific cues present in high frequency zone is proposed in \cite{Sandipanetal2003}. Both of them were combined to improve the performance of the MFCC-based system. A multilevel speaker recognition system was proposed in \cite{Joaquinetal2007} combining acoustic, phonotactic and prosodic subsystems. Acoustic features used were the MFCCs. Speaker recognition system in noisy environments was improved by use of a SNR sensitive subspace based enhancement technique and probabilistic SVMs \cite{Wangetal2007}. In this paper MFCCs were used to represent the speech features. Error rate for identification in noisy environments was reduced from 43.4\% to 25\%.\par The speaker verification system presented in \cite{Wilsonetal2006} uses the following set of features: the first four formants, the amount of periodic and aperiodic energy in the speech signal, the spectral slope of the signal and the difference between the strength of the first and second harmonics. The NIST ’98 Evaluation Database was used for the experiments which consisted of telephone speech sampled at 8 kHz. The speaker models were constructed using the GMMs and were trained using maximum-likelihood parameter estimation. For female speakers the error rate was found to be 36.69\% and for male speakers the error rate was found to be 34.4\%. The features performance was better compared to MFCCs for female speakers but not for male speakers.\par Another set of features were presented in \cite{Beraneketal2003} for speaker verification. The parameters considered were fundamental frequency, the articulator configuration of nasal consonants captured by the amplitude normalized filter outputs in the nasal spectra, the first two formants of the vowels, source spectrum slope and prevoicing parameters. Speaker verification was done using neural networks in \cite{Templetonetal2003}. The neural networks were trained with unweighted cepstral coefficients derived from Linear Predictive Coding (LPC). The convergence method used in the network is the back-propagation learning algorithm of Multi-Layer Perceptron (MLP).\par In this paper we propose four simple intrapitch temporal features: positive crest, positive trough, negative crest and negative trough in conjunction with pitch synchronous cepstral coefficients. These features extracted from the steady state vowel region is used to characterize a speaker. The speaker recognition is possible mainly due to the vibration of the vocal folds as mentioned in \cite{Pradhanetal2011}. Hence the proposed features are extracted from steady state region of the utterances i.e. the vowel region. \par The remaining of the paper is organized as follows: Section 2 describes the basic methodology and Section 3 describes the experiments and results and Section 4 describes the conclusions and future work.
\end{justify}

\section{METHODOLOGY}
\begin{justify}
The basic methodology followed in this paper can be divided into two phases namely the “preprocessing, pitch detection and marking phase” and “the processing phase”. In the preprocessing phase, the pitch period is first detected and marked. In the processing phase the proposed features are computed from the steady state regions of the vowel. This processing is applied to both the training data and the testing data. Initially the system is trained with 20 speakers. In the preprocessing phase for each speaker, pitch periods are detected by marking the position of the start of each pitch cycle. The proposed features and pitch synchronous cepstral coefficients from each of the pitch periods in the speech signal is calculated in the processing phase. The average of all 20 utterances gives the model for a given speaker. In the testing phase, the pitch is detected and the features are extracted from the test sample and the distance of the features set from each of the 20 models is computed using Tokhura’s distance. The speaker model for which both the cepstral coefficients and the proposed feature set gives the minimum distance is declared as the recognized speaker.
\end{justify}

\subsection{PREPROCESSING PHASE}
\begin{justify}
The speech signal was recorded using the Cool Edit software and saved in the text format. This file (input signal) is used for all processing. Initially the DC shift correction is applied to remove the DC components that might have got added during the recording or due to power supply interference. Next, the samples are normalized to a value depending on the sampling rate of the signal (in experiments normalization value is fixed at 10,000 after some tests). Silence removal is the next step in the preprocessing phase where the average energy of the non speech activity signal is calculated and removed those frames whose energy is less than a given threshold. In the experiments a frame is classified as a speech frame if the average energy is beyond 110\% of the average silence energy. The size of the frame considered is 100 samples shifted by 50 samples for subsequent frames. The main step in the preprocessing phase is pitch detection. Pitch is detected by the following algorithm:\begin{enumerate}
\item Two files are maintained. One containing the peak sample value in each of the positive halves along with the maximum of the difference of that sample from the previous half and next half, say Maximum Peak Difference (MPD), (halves should be compatible i.e. if current peak is in the positive half, the previous and the next half considered should be positive)  and the second containing the peak sample value in each of the negative halves (maintained in the second file) along with its MPD values in the negative halves.
\item From the sample values of the above mentioned two files, average of the differences is computed (i.e. Average of the Maximum Peak Differences (AMPV)) for positive half as well as negative half. The pitch is calculated with respect to the positive half or the negative half based on consistency.
\item Now based on the result of the above computation, either the positive half or negative half is used for pitch calculation and the pitch period is calculated as:
$$Pitch period=Index(S_{T}) - Index(P_{T})$$
$$Threshold = Current_{peak value} - (\frac{x}{100} *  Current_{peak value})$$\\
where $S_{T}$ is Sample having value greater than or equal to Threshold, $P_{T}$ is Previous Threshold, x is determined based on the MPD of the current peak sample value. If the MPD is less than AMPV then AMPV is divided into 10 intervals and the interval in which MPD falls is found out and the interval number is taken as the value of x (i.e. if MPD falls in the interval 5 then value of x is 5). Similarly if MPD is greater than AMPV then difference between the maximum (all the MPDs) and AMPV is divided into an interval of 10 and the interval in which MPD falls is found out and the value of x is set as 10+x (i.e. if the MPD falls in the $3^{rd}$ interval then x is set as 13). 
\end{enumerate}
The pitch period obtained will be further used in feature extraction. The start and end of each pitch period is marked and kept for feature extraction.
\end{justify}

\subsection{PROCESSING PHASE}
\begin{justify}
Now using the pitch information computed from the Preprocessing phase the features are extracted. The features considered are positive crest, positive trough, negative crest and negative trough and is shown in Figure \ref{Featureset}.

\begin{figure}[H]
\centering
\includegraphics[width=1\textwidth]{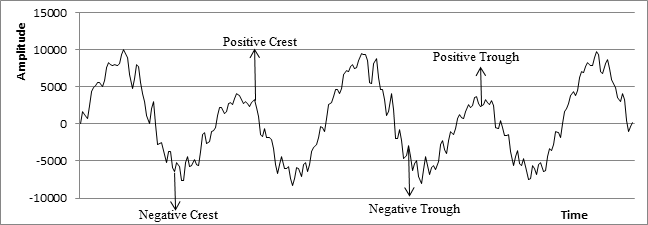}
\caption{Feature set computation of the vowel /i/.}
\label{Featureset}
\end{figure}

\par The feature extraction algorithm is as follows: \begin{itemize}
\item For feature extraction, the steady state region of the vowel is identified. For this, maximum 10 frames before and 10 frames after the frame containing the normalized value are considered. These frames are steady state regions of the utterances (i.e. vowel part of the word). Each frame is of length equal to pitch period as computed above. Sometimes depending on the speakers utterance 10 frames may not be available. So in that case, available number of frames is taken and the features are calculated.\item Now a window of three speech samples is considered and slide the window through the pitch period. Check whether the middle value of the window is greater than the other two (by magnitude), if true, increment the respective counter (if it is positive half increment the crest counter else increment the trough counter) and advance the window by one sample. If the middle value of the window is lesser than the other two values (by magnitude) then increment the respective counter (for positive half trough counter is incremented and for negative half crest counter is incremented) and advance the window by one sample. Else just advance the window by one sample. \par Suppose \textit{poc} is the counter for positive crest, \textit{pot} is the counter for positive trough, \textit{nec} is the counter for negative crest and \textit{net} is the counter for negative trough and w[3] represents the window of processing, then the computation of the features can be shown mathematically as follows:

\begin{itemize}
       \item[]If w[2]$>$0\begin{itemize}
	\item[]If w[1]$<$w[2] and w[2]$>$w[3]\begin{itemize}\item[]$poc=poc+1$ and advance the window by one sample\end{itemize}
	\item[]Else if w[1]$>$w[2] and w[2]$<$w[3]\begin{itemize}\item[]$pot=pot+1$ and advance the window by one sample\end{itemize}
	\item[]Else\begin{itemize}\item[]advance the window by one sample\end{itemize}\end{itemize}
      Else\begin{itemize}
	\item[]If w[1]$<$w[2] and w[2]$>$w[3]\begin{itemize}\item[]$nec=nec+1$ and advance the window by one sample\end{itemize}
	\item[]Else if w [1]$>$w[2] and w[2]$<$w[3]\begin{itemize}\item[]$net=net+1$ and advance the window by one sample\end{itemize}
	\item[]Else\begin{itemize}\item[]advance the window by one sample.\end{itemize}\end{itemize}
\end{itemize}
	
\item The features are computed over the 20 frames (if available) of an utterance and the average is computed. If N is the number of frames available, the final feature value for an utterance will be
$poc=\frac{poc}{N}$\hspace{8pt}$pot=\frac{pot}{N}$\hspace{8pt}$nec=\frac{nec}{N}$\hspace{8pt}$net=\frac{net}{N}$ \item Standard cepstral coefficients (computed using the Durbin’s algorithm) are also used along with the above described feature set. Thus a total of 16 features are used for the speaker recognition system (12 cepstral coefficients and 4 proposed features).\item Cepstral coefficients are calculated over the frame F1 of length equal to three consecutive pitch periods say f1, f2, f3 (within the previously defined range of 10 frames before and after the normalized frame) i.e. $F1=f1+f2+f3$.
\item In each iteration, the frame is shifted by a value equal to f1 (i.e. $F1-f1$) to the right and pitch of the next frame is added (i.e. $F2=F1-f1+f4$) and cepstral coefficients are calculated (thus a maximum of 18 iterations are available in cepstral coefficient calculation).
\item The average of the above said 18 frames are calculated and it is the cepstral representative of a particular vowel utterance.
\item The average of the feature set (16 values) over the 20 utterances of each speaker is computed. This average value serves as the model for that speaker.\end{itemize}
\end{justify}

\section{EXPERIMENTS AND RESULTS}
\begin{justify}
The experiments consist of training the system with steady state vowels from a set of speakers and testing the validity or claim from an unknown vowel utterance of an unknown speaker. The speaker verification system was developed using Microsoft Visual C++ and it was trained with 2000 utterances taken from 20 male speakers in the age group of 22-26 years. The following five sentences were used for recording and the vowel regions were extracted manually and the preprocessing and the processing were applied on the extracted vowel regions.\begin{enumerate} \item An age of apes is an old story.(/e/)\item There is a river by the mango tree.(/o/)\item Ramayana is an old story.(/a/)\item A cuckoo sings on a mango tree by the river.(/u/)\item The bee is sitting on a flower by the mango tree.(/i/) \end{enumerate}\par The steady state part of the five cardinal English vowels were marked and saved for feature analysis. The quality of the accuracy of the system was calculated based on another set of 500 utterances from the same set of 20 speakers. During the training phase, the model (feature set) of each speaker was determined by averaging the 20 training utterances of each vowel. During the testing phase, test samples were collected and the feature set was extracted from the samples and distance of the feature set from all the models (with which the system is trained) was calculated. The unknown speaker was identified to be the speaker $S_{i}$ with which cepstral coefficients as well as the new feature set (of the test sample) had minimum distance. If the cepstral coefficients was found to be showing minimum distance with speaker $S_{i}$ and new feature set was found to be showing minimum distance with speaker $S_{j}$ ($i \neq j\hspace{2pt}\&\hspace{2pt}1 \leq i, j \leq 20$ )
then it is considered to be an invalid case and was rejected. In such cases, the speaker is told to speak once more. Even though the rejection rate is high, in the case where it is valid, the new system is much more accurate than the standard cepstral-based speaker recognition system. The accuracy of the new feature set based speaker recognition system is low compared to cepstral based speaker recognition system. But the combined system is much more accurate compared to cepstral based recognition system. The accuracy of the cepstral based system, feature based system and the combined system is shown in the Table \ref{AccuracySystems} below:\end{justify}

\begin{table}[H]
\begin{tabular}{|c|c|c|c|}
\hline
Parameters & Total Test Cases & Correctly Recognized & Accuracy\\
\hline
Cepstral coefficients & 500 & 347(out of 500) & 69.81\%\\
\hline
New Feature Set & 500 & 144(out of 500) & 28.97\%\\
\hline
Combined & 500 & 122(out of 134) & 91.04\%\\
\hline
\end{tabular}
\caption{Comparison of the accuracy of the various systems.}
\label{AccuracySystems}
\end{table}

\begin{justify}
\par The accuracy of the speaker verification system for each vowel is given in the Table \ref{AccuracyVowels} below:\\
\end{justify}

\begin{table}[H]
\begin{tabular}{|c|c|c|rl|}
\hline
Vowel & Total Utterances & Rejected Cases & Accepted & cases \\\cline{4-5} & & & \multicolumn{1}{|c|}{Correct} & Wrong\\
\hline
/a/ & 100 & 77 & 17(73.91\%) & \multicolumn{1}{|c|}{6}\\
\hline
/e/ & 100 & 66 & 33(97.05\%) & \multicolumn{1}{|c|}{1}\\
\hline
/o/ & 100 & 77 & 19(82.60\%) & \multicolumn{1}{|c|}{4}\\
\hline
/u/ & 100 & 79 & 21(100\%) & \multicolumn{1}{|c|}{0}\\
\hline
/i/ & 100 & 67 & 32(96.97\%) & \multicolumn{1}{|c|}{1}\\
\hline
\end{tabular}
\caption{Accuracy of the individual vowels in speaker verification with the proposed feature set.}
\label{AccuracyVowels}
\end{table}

\begin{figure}[H]
\centering
\includegraphics[width=1\textwidth]{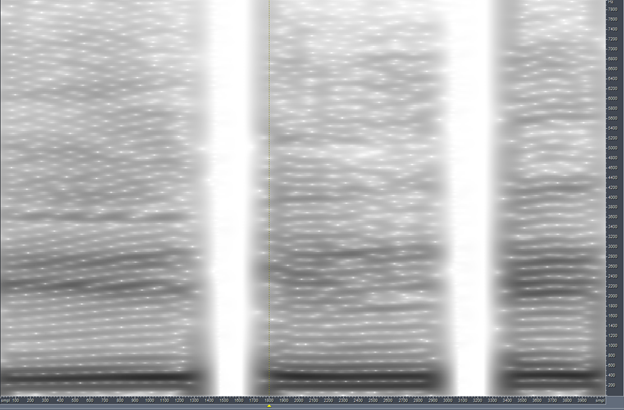}
\caption{Spectrogram of the speaker with which mismatching occurred ,correct speaker and wrongly recognized test speaker}
\label{Spectrogram}
\end{figure}

\begin{justify}
\par The selection of the words from which the vowels were extracted had a significant contribution in the accuracy of the speaker recognition system mainly for the vowel /i/. Initially /i/ was extracted from the word “river”. In that case because of the short duration of /i/ and the preceding consonant, the vowel part was smoothened. So for vowel /i/ the word used was “bee” where the duration of sound /i/ was sufficient and the waveform was also consistent. \par There were some utterances where the cepstral based system showed misrecognition was correctly recognized by the new feature set. It was also seen that some of the utterances where new feature set showed misrecognition was correctly recognized by the cepstral based system. In certain cases both cepstral and new feature set showed misrecognition. All these cases were considered as unreliable cases where the decision could not be made. Such samples were rejected by the system. Even though the rejection rate is more, the decision made by the system in reliable cases is much more accurate compared to the cepstral based system and the new feature set when considered alone.  \par In the misrecognised cases, when the graphs were plotted with the feature set it was found that the misrecognition had occurred for those speakers whose utterances (from the training set) varied a lot. The graph corresponding to the feature set of the misrecognized speaker is shown in Figure \ref{VowelGraph}.

\begin{figure}[H]
\centering
\includegraphics[width=1\textwidth]{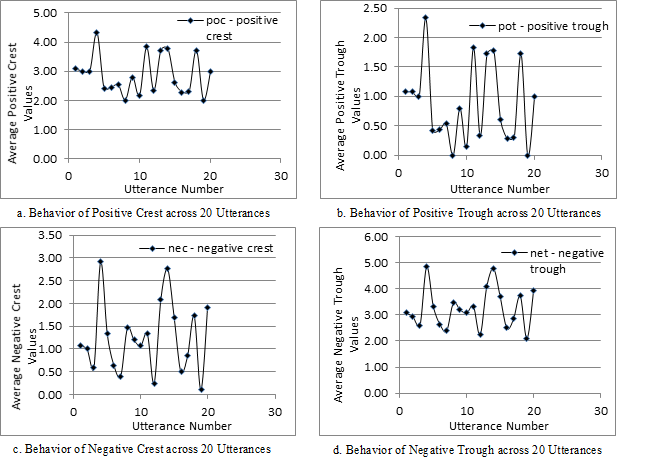}
\caption{Graph of the behavior of the four features across the 20 utterances of a misrecognized speaker for vowel /u/}
\label{VowelGraph}
\end{figure}

\par Here the variation throughout the utterances can be seen as a non uniform distribution of the feature set values. So recognition in such speakers is low as the feature set values does not concentrate to a specific value and it can get matched to any other speaker. It was also noted that for the vowel /u/ and /o/ even a slight variation in the feature set can lead to misrecognition (because of the narrow gap between the feature set values). In this case the number of misrecognition is limited by the cepstral coefficients. But for vowel /i/ the acceptance range of variation in feature set is more compared to all other vowels. This is because of the wide difference in the feature set values for the speakers. \par Speech spectrogram of the speaker with which a test sample was wrongly recognized, the correct speaker it should have been matched with and the test sample itself is shown in Figure \ref{Spectrogram}.
\end{justify}

\section{CONCLUSIONS}
\begin{justify}
In this paper, we have proposed a set of features which can be used to characterize speakers. The feature set consists of four simple intrapitch temporal attributes: positive crest, positive trough, negative crest and negative trough along with 12 pitch synchronous cepstral coefficients. The system was tested with a set of 20 speakers. The accuracy of the system was found to be 91.04\%. The vowels /i/, /u/ and /e/ were found to be more accurate in speaker verification system. The future work may focus on the automation of the separation of vowel region from other regions of the recorded sentences and addition of more features to improve the accuracy. Experiments are being done with more speakers to check the change in accuracy of the system.
\end{justify}

\bibliography{Speakerverification}
\end{document}